\documentstyle [fleqn,epsf]{article}
\begin{document}


\newcommand{\inplot}[2]{
 \begin{center}
 \leavevmode\epsfysize=#2mm \epsfbox{#1}
 \end{center}
}

\newcommand{\va}{\vspace{4mm}}
\newcommand{\vb}{\vspace{8mm}}
\newcommand{\vs}{\vspace{8mm}}

\newcommand{\be}{\begin{equation}}
\newcommand{\ee}{\end{equation}}
\newcommand{\ba}{\begin{eqnarray}}
\newcommand{\ea}{\end{eqnarray}}
\newcommand{\nns}{\nonumber \\}
\newcommand{\nn}{\nonumber}
\newcommand{\NL}{\nonumber \\}

\newcommand{\ra}{\longrightarrow}

\newcommand{\pd}[2]{ \frac{\partial #1}{\partial #2} }


\begin{Large}                                


{\bf Towards the Equation of State of Classical SU(2) Lattice }

\centerline{\bf Gauge Theory}

\end{Large}

\centerline{{\sc \'A. F\"ul\"op, T. S. Bir\'o}}

\centerline{RMKI, KFKI}

\centerline{H-1525 Budapest, P.O. Box 49, Hungary}



\va


\vspace{0.8cm}

{ \bf Abstract}

\vspace{0.4cm}

We determine numerically  the full complex Lyapunov spectrum of SU(2) 
Yang-Mills fields on a 3-dimensional lattice
from the classical chaotic dynamics 
using eigenvalues of the mo\-no\-dro\-my matrix. 
The microcanonical equation of state is determined as the entropy
-- energy relation utilizing the Kolmogorov-Sinai entropy
extrapolated to the large size limit.


\vspace{0.8cm}

{\bf Introduction}

\vspace{0.4cm}

Knowledge of the microscopic mechanism responsible for the local 
equilibration of energy and momentum in the nuclear 
collision is fundamental to understand the possible 
emergence of a quark gluon plasma in relativistic heavy-ion collisions. 
It is evenly essential to study the equation of state of this system
(in and out of equilibrium) in order to be able to make
predictions or interpret experimental data from the perspective
of quark -- gluon matter.

\va
Many of the approaches to the equation of state assume
thermal equilibrium, since the real -- time evolution of such a
plasma in the full quantum field theory cannot yet be followed
in a non-perturbative manner by presently practized numerical
methods. Perturbative QCD helps to get hints from the high-energy
parton dynamics, but the non-perturbative lattice gauge theory
is difficult to extend to non-equilibrium phenomena.

\va
Fortunately, the situation is eased by findings which point to
the possibility that the quark -- gluon plasma can be dynamically
studied in a semiclassical approximation. Hard thermal loop
resummation has been shown to cope with the semiclassical transport
(linear response) theory approach \cite{blaiz1,blaiz2}. 
In fact earlier numerical studies of the classical Hamiltonian dynamics of
Yang-Mills systems on three -- dimensional lattices revealed
an intriguing coincidence between the maximal Lyapunov exponent
and twice the gluon damping rate \cite{bt}. 
In the background of that argumentation equipartition
of the energy has been conjectured as a result of the classical
chaotic dynamics on the lattice. This conjecture has been
supported by the pseudo -- Boltzmannian distribution of
one-plaquette energies (the local magnetic energy density) in
long-evolved classical configurations.

\va
The correspondence between this average energy and the
Kolmogorov-Sinai entropy has been first investigated in \cite{CG} 
for pure SU(2) Yang-Mills systems. Those times, however, the
Lyapunov spectrum could be obtained only for relatively small
systems (N=2,3) with the rescaling method. This way also only
the positive real exponents could have been calculated.
The question of the thermodynamical limit ($N \ra \infty$)
and the investigation of a possible dependence on the
initial configuration (which is usually done by using improved 
statistics) remained open. (Citation: ``Future work is still
required here'' [\cite{CG}, page 6].)

\va
In the present article we study the ergodization of SU(2)
lattice gauge theory due to its classical chaotic dynamics
by using different methods as before. We also consider
larger lattices ($N=2,3,4,5,6$) for obtaining the entropy
and extrapolate to the large $N$ limit. In our approach
the {\em full complex spectrum} of Lyapunov exponents
will be obtained using the mo\-no\-dro\-my matrix 
(linear stability matrix
along the path determined by the classical equation of motion)
in the space of evolving field configurations on the lattice.
The complex spectrum offers an insight also into the periodic
(oscillatory) behavior of field fluctuations; a basic experience
for learning about the corresponding quantum theory.

\va
The question of ergodization is addressed via the Kolmogorov-Sinai
entropy, which is calculated up to linear lattice sizes of
$N=6$ (using $24N^3=5184$ dimensional phase-space) and extrapolated
to $N \ra \infty$. We extrapolate both for long times, --
in order to get close to the Lyapunov exponents, -- and for
large systems -- in order to approach the thermodynamical limit.
Ensembles of randomly chosen (i.e. chosen according to the
Haar measure in the group variables) initial configurations
with a given (slightly fluctuating) total energy are averaged
over, too.

\va
This article is constructed as follows. First we review basic definitions
dealing with the chaotic behavior of extended systems and then
the Hamiltonian treatment of classical lattice gauge theory.
Finally we present results on full complex Lyapunov spectra,
scaling properties, extrapolations and the entropy - energy
relation (microcanonical equation of state) for SU(2) lattice
gauge theory.


\vspace{0.8cm}

{\bf Measuring Chaos}

\vspace{0.8cm}

Instead of the classical determination of the Lyapunov exponent, 
which has been calculated by the rescaling 
method in \cite{CG,bmat,cgay,tsbagbmat,bffm}, 
in this article we  use the mo\-no\-dro\-my matrix 
approach. The Lyapunov spectrum $L_i$  is expressed in
 terms of the mo\-no\-dromy matrix's eigenvalues $\Lambda_i$:
\be
L_i = \lim_{T \rightarrow \infty} 
\frac{\int_0^T \Lambda_i(t)dt}{T} \;\;\;\;\;\;\; i=1....f
\label{LYAPU}
\ee
where the $\Lambda_i(t)$-s are the solutions of the
characteristic equation 
\be 
{\rm det} \left( \Lambda_i(t) 1 - M(t) \right) = 0  
\label{CHAR}
\ee
at a given time $t$.
Here $M$ is the linear stability matrix,  
$f$  is the number of degrees of freedom. 

\va
We consider   conservative dynamics   
fulfilling  Liou\-ville's theorem. 
\be
\sum_{i=0}^f  L_i =0  
\label{LIOUVILLE}
\ee

In the numerical calculation we use the discrete definition of the 
Lyapunov spectrum:

\be
L_i^{'}=   
 \left< \Lambda_i \right>^{(n)} = \frac{1}{n} \sum_{j=1}^n \Lambda_i(t_{j-1}) \\
i=1...f, 
\label{SERIES} 
\ee
where $t_j$ are subsequent times along an evolutionary path of the 
gauge field configurations.  
Nevertheless a few initial configurations has been omitted from
the average, although the starting configurations has been
produced randomly.
We  need to extrapolate the quantities $L_i^{'} $ to the 
long time limit $ n \rightarrow \infty, $  with fixed timestep.
It is expected to converge to Lyapunov exponent 
$L_i$ occuring in eq. (\ref{LYAPU}) for a non-compact
configuration space. As we shall see by discussing the results of 
numerical simulations gauge group systems live in a compact
configuration space, therefore this limit is not entirely
safe in our case. The rescaling method observes pairs of gauge
field configurations which are close in the phase space, so
it has a control on the distance of these two and can monitor 
this quantity not growing close to or over
the limiting distance given by the compact size of the phase
space. It is achieved  by doing frequent rescalings. 
The monodromy matrix method on the other hand follows
only one gauge field evolution, in this case the short and long time
behavior can be different.

\va
The Kolmogorov-Sinai entropy is obtained  by using  Perin's formula:
\be
h^{KS}= \sum_i L_i \Theta(L_i) 
\ee
using the theta function $\Theta(x)$ being one for positive arguments and
zero otherwide.
The dimension of $h^{KS}$ yet is a rate (1/time), so for estimating
the entropy we shall use the normalized quantity:
\be
 S=\frac{h^{KS}}{Re(L_0)N^3}.
\ee
The redefinition is done in a way typical for extensive
quantities, here the entropy density, on an $N\times N\times N$ lattice.



\vs
{\bf Lattice Yang-Mills field systems}

\vs

Our discussion is based on the Hamiltonian formulation of 
the classical lattice SU(2) gauge theory. It is
governed by the Hamiltonian \cite{BOOK}:
\be
H=\sum_{x,i}\left( \frac{1}{2} \left< \dot{U}_{x,i} , \dot{U}_{x,i} \right>
 +  \left( 1-\frac{1}{4}\left<U_{x,i}, V_{x,i}\right> 
\right)\right)
\ee
where $U_{x,i}$ is the SU(2) group element lying on  an oriented  
link of the lattice  starting at the 
site x,  pointing to the i  direction.  
In the numerical rea\-li\-zation it is represented by real
quaternions.  The notation $<A,B>$ belongs to the normalized 
trace of the product of two SU(2)
group elements : 
\be
<A,B>=\frac{1}{2}tr(AB^{-1}).
\ee  
The $V_{x,i}$  complement matrix is  constructed from triple products of 
link variables, which complete  the
considered link (x,i) to an elementary plaquette.
\be
 V_{x,i}= \frac{1}{4} \sum_{i,j,k}U_i U_j^{-1} U_k^{-1} 
\ee
Finally the overdots denote the derivative with respect to the
scaled time, $t/a$, as well as the Hamiltonian $H$ stands for
the scaled energy $g^2aE$.

\vs
The Hamiltonian equations of motions are given by 
(suppressing the link indices in the rotation):
\ba
 \dot{U} & = & P, \NL
 \dot{P} & = & V - \left< U,V \right> U -\left <P,P \right> U. 
\ea
These equations of motion conserve unitarity $<U,U>=1$ , orthogonality $<U,P> = 0$, 
and Gauss' law.  The mo\-no\-dro\-my matrix is defined as:
\be
M = \left(\begin{array}{cc} 
    \pd{\dot{U}}{U} &
    \pd{\dot{U}}{P} \\
 \\
    \pd{\dot{P}}{U} &
    \pd{\dot{P}}{P} \\
 \end{array} \right)
\ee 
Here the different partial derivatives are obtained by comparing the above
equations of motion for the configurations $(U,P)$ and 
$(U+\delta U, P+\delta P).$ We arrive at
\ba
 \pd{\dot{U}^a}{U^b} &=& 0, \NL
 \pd{\dot{U}^a}{P^b} &=& \delta^{ab}, \NL
 \pd{\dot{P}^a}{U^b} &=& \pd{V^a}{U^b} - \left(U_c \pd{V^c}{U^b} \right)
   U^a - V^bU^a - (U_cV^c + P_cP^c)\delta^{ab}, \NL
 \pd{\dot{P}^a}{P^b} &=& -2P^bU^a.
\ea
These expressions provide  information about the stability of 
trajectories in the neighbourhood of 
any point of an  orbit in the (U,P) phase space. A small  perturbation, 
\hbox{$( \delta U, \delta P )$}
evolves in time governed by the monodromy matrix $M.$ 
The eigenvalues of this matrix can be classified as follows: 
for real and positive eigeinvalues neighbouring trajectories apart 
exponentially and the motion is unstable. In the limit of large time
we obtain the Lyapunov exponents from these eigenvalues. 
The imaginary parts of the complex eigenvalues describe
oscillatory frequencies of perturbations. 

\vs
In contrast to the rescaling method, where only 
the positive real part of the Lyapunov exponent can be measured,
the study of the spectrum of the mo\-no\-dro\-my matrix makes 
these complex eigenvalues 
numerically available for the first time. The total number of 
degrees of freedom in the numerical
simulation is given by $f=4*3*N^3=12*N^3$, using four-real element 
quaternions for  SU(2) group elements (so the phase space is
$2f=24N^3$ dimensional). Due
to the constrains $<U,U>=1$, however, the physically relevant number of 
degrees of freedom is the same as in earlier simulations, 
which used $E^a$, and ${\rm tr}(U\tau^a)$ as independent degrees of freedom.


\vs

{\bf The spectrum of the stability matrix}

\vs

The eigenvalue spectrum $\Lambda_i$ of the mo\-no\-dro\-my matrix
consists of a  real subset and a complex subset on the complex plane 
(cf. Figs. 1 and 2).
Although this $2f \times 2f$ sized matrix is sparse, since each
group element exerts a force onto close links (those sharing a
common plaquette) only, the usual sparse matrix methods are not
applicable, because they determine just a few leading eigenvalues.
In order to determine all complex eigenvalues with an acceptable
precision, only `brute force' methods with a memory need of
${\cal O}(f^2)$ and computational time of ${\cal O}(f^3)$
can be used. The maximal system size, our computational resources
allow us to use, belongs to $N=6,$ meaning a $2f=5904$
dimensional phase space.

\vs
Figs.1 and 2 display the complex eigenvalues for several configurations
during the evolution along a few trajectories. At high energy
per degree of freedom, $g^2aE=0.8,$ (near to the saturation value $1$)
the region covered by the eigenvalues in the complex plane
shows a mirrorred ``butterfly'' shape, while at low energy,
$g^2aE=0.1,$ rather two ``ovals''. In all cases the covered region
is symmetric both to the real and imaginary axes: the former property
is due to the fact that the equations of motion are real, the latter
is due to the Hamiltonian is conservative (time independent).
It is interesting to observe that about one sixth of the
eigenvalues are purely real, not allowing for oscillations
or wave-propagation in the fluctuations.
In neither case shows the eigenvalue region any resemblence
of Wigner semicircles, like those found in studies of the
eigenvalues of the Dirac operator in random gauge field background
\cite{SEMI},
and usually regarded as a signature of quantum chaos.
Here we deal with a bosonic system at comparatively high
excitations and the chaotic behavior is quite classical.

\vs
According to the imaginary part of the eigenvalues the
pattern separates to three islands at all energies.
An important qualitative feature of these patterns is
the gap between zero and non-zero imaginary parts:
it behaves as a dynamically developed infrared cut-off
(``gluon mass'') for oscillations in small perturbations
of the classical equations of motion. Also a number of
zero-frequency modes occur, this is connected to
symmetry transformations commuting (in the Poisson bracket
sense) with the Hamiltonian, such as time independent gauge 
transformations. The gap seems to be reduced and eventually 
disappear at high energy. Do these differences between low and 
high energy establish a two-phase picture of the classical
lattice SU(2) system? In order to answer this question
a study of the equation of state is necessary.


\vs
{\bf Scaling and Equation of State}
\vs

We aim to obtain the equation of state from the dynamical
simulation. For this purpose we first study the finite
size scaling and extrapolate to infinite ($1/N=0$) lattices.
Then we study the energy dependence of the maximal
Lyapunov exponent as well as the Kolmogorov-Sinai entropy.
The latter leads eventually to the equation of state
as the entropy - energy relation, $S(E),$ in the
thermodynamical limit of infinite volume.

\vs
Fig.3 shows the real part of the  Lyapunov spectrum
extrapolated to $1/N \rightarrow 0$ from data taken at
$N=2,3,4,5$ and $6$ at high energy ($g^2aE=0.8$).
The overall pattern is similar to that obtained earlier for
smaller systems ($N=2,3$), just the number of purely
imaginary eigenvalues (for those ${\rm Re}L_i = 0$)
is greater due to our use of more variables (and, of course,
more constraints). The structure of the ordered real part
of the Lyapunov spectrum is similar at all energies, but the
maximal point, $aL_0,$ scales with the energy, $g^2aE.$

\vs
Fig.4 displays the fluctuation of the maximal real part
Lyapunov exponent, $L_0,$ originating in different, randomly
chosen initial field configurations. This levels at a few
per cent. The energy scaling of the maximal Lyapunov
exponent has been debated in the past and has been found
linear in the long-time limit.
Doubts rose for low energies \cite{nrr,nrr1}, stating that the
correct scaling here would be $L_0 \sim E^{1/4}.$
More elaborated studies with the rescaling -- aparting
method showed then a tendency back to the linear,
$L_0 \sim E$ scaling also at low energies for long enough
time observations \cite{bm,hhlmm,jbbmas}.
Our data agree with the linear scaling with a coefficient
of $0.5$ for a middle to short time evolution, as it can be
seen in Fig.5. Surprisingly, and not yet fully understood,
the very long time behavior of the maximal real part of the
monodromy matrix eigenvalues show a definitely sublinear
scaling with the energy per degree of freedom. The best fit
is, however, not like $E^{1/4},$ but rather a logarithm
(cf. Fig.6). We suspect that following a trajectory too long
makes the observed eigenvalues feel the compactness of
the configuration space -- which is an artefact of the
lattice field theory Hamiltonian. In the following discussion
we refer to data with linear energy scaling only.

\vs
Fig.7 displays the extrapolation of the maximal Lyapunov
exponents to the thermodynamical limit at different
energies. The correspondence proved to be almost linear
by assuming an
$$ L_0 \sim \frac{1}{\sqrt{f}} \sim N^{-3/2} $$
scaling with the finite size. This corresponds to sampling
ergodic states \cite{jbbmas}. In particular at high energies the
extrapolated $L_0$ is higher than any actually obtained
value in simulations at the finite $N.$

\vs
Finally we obtained the Kolmogorov-Sinai entropy from
infinite-size extrapolated, initial state and evolution
averaged eigenvalue spectra as a function of the scaled
energy. For a (nearly) ideal gas on expects an
$S \sim \log E$ relation, indeed this is a good
approximation of our data (cf. Fig.8). We obtain
\be
\langle S \rangle = (0.5084 \pm 0.023 ) \log (g^2aE)
\, + \, (2.3334 \pm 0.0452).
\ee
The indicated errors after the $\pm$ signs are of
statistical nature. This best fit belongs to an inverse
temperature
\be
\frac{1}{T} = \frac{\partial \langle S \rangle }{\partial E}
\approx \frac{0.5}{E},
\ee
describing equipartion with
\be
 E = \frac{1}{2} T.
\ee 
Based upon our calculations even a two-phase structure
(or a crossover) cannot be excluded with absolute safety:
at mid energies a depletion is hinted in the data.
The $S(E)$ curve would show a first order two-phase
structure by having a break somewhere. For a possible
relevance to lattice SU(2) systems see the paper \cite{s}.
In our case, starting from the chaotic dynamics of
classical lattice Hamiltonians, a refined study is needed
in the transition region $g^2aE = 0.2, \ldots 0.5$ in order
to settle this question.


 
 
  
\vs
 
{\bf Acknowledgements}

\vs

One of us (\'A.F.) gratefully acknowledges 
discussions with professor Tam\'as T\'el.
This work has been supported by the Hungarian -- American
Joint Fund MAKA T\'eT (JF.Nr. 649).

\vs

\vs


\newpage


{\Large \bf FIGURES}
\vb

$\qquad$
\vs

\begin{figure}[h]
\inplot{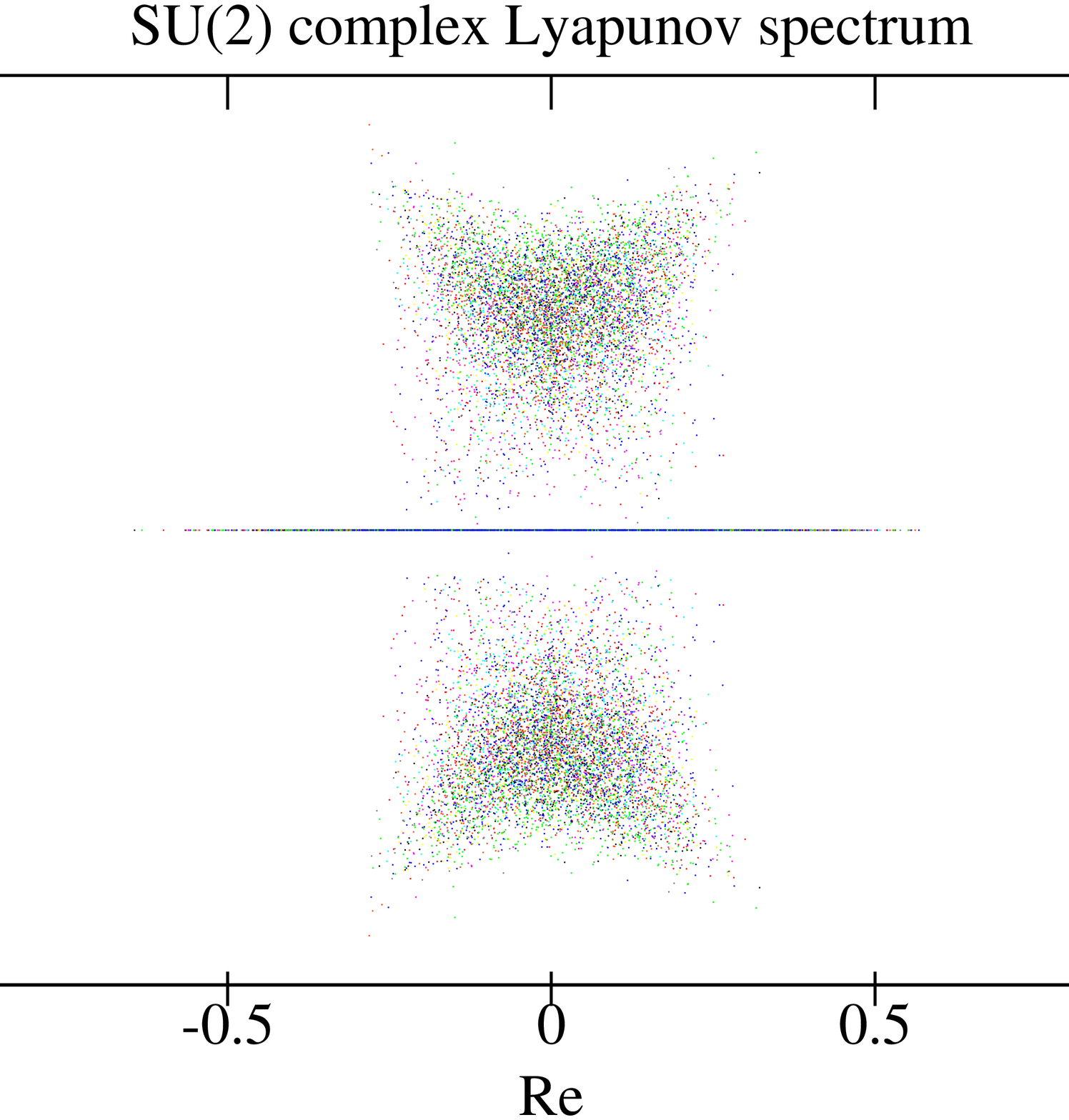}{85}
\caption{ 
 Full complex eigenvalue spectrum of monodromy matrix: high energy.  
}
\end{figure}

\begin{figure}
\inplot{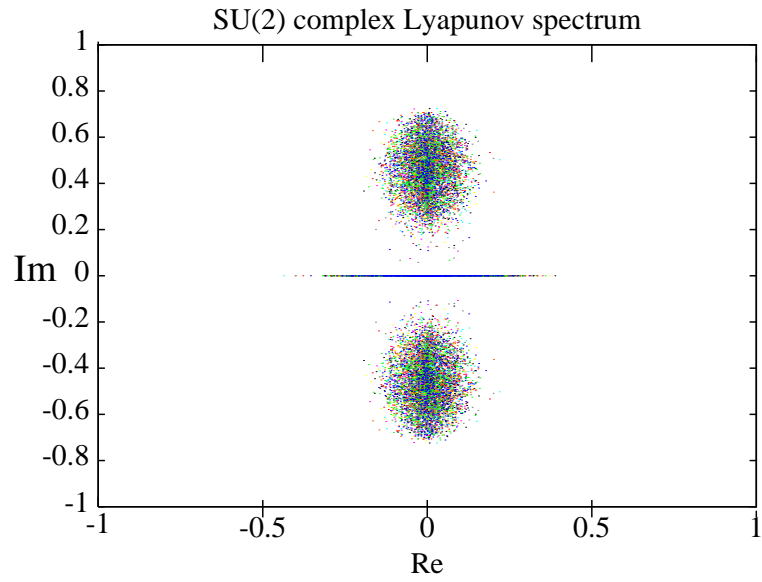}{85}
\caption{
 Full eigenvalue spectrum of the mo\-no\-dro\-my matrix: low energy.
}
\end{figure}

\begin{figure}
\inplot{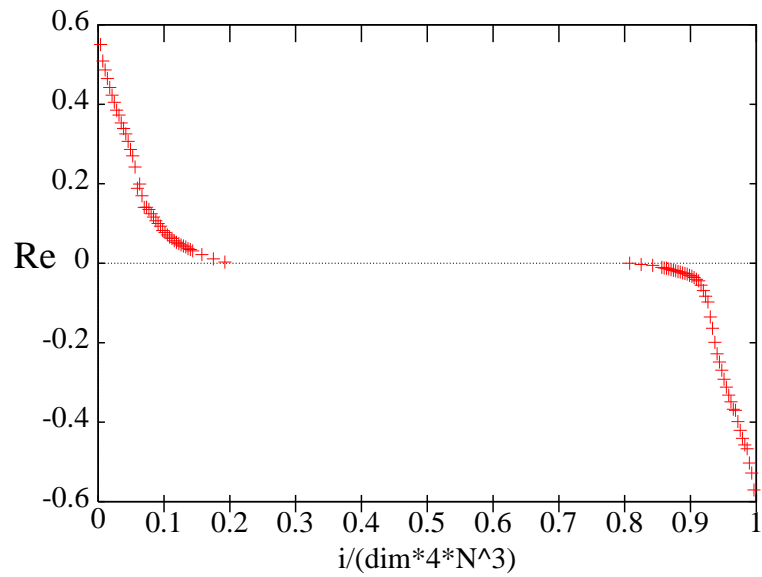}{85}
\caption{
 The real part of the Lyapunov spectra in the case of $N \rightarrow \infty$
}
\end{figure}

\begin{figure}
\inplot{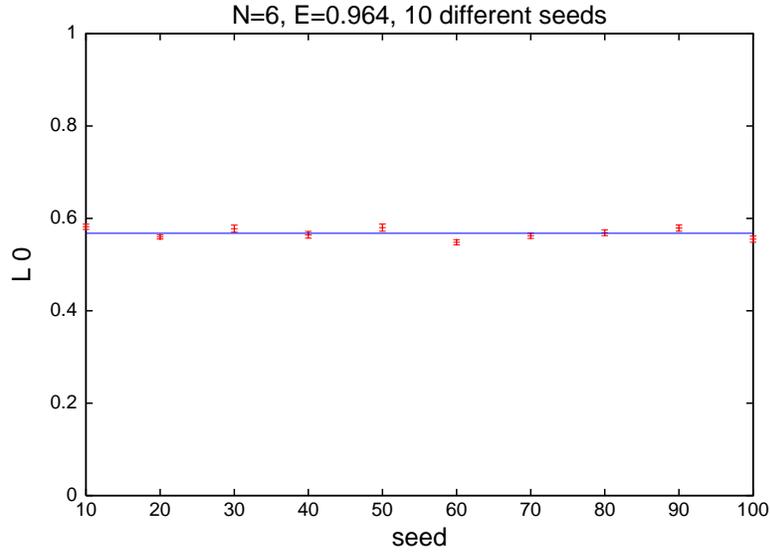}{85}
\caption{
 Maximal Lyapunov exponent obtained starting from different
 random configurations on an $N=6$ lattice at scaled energy 
 $g^2aE = 0.964$. 
}
\end{figure}

\begin{figure}
\inplot{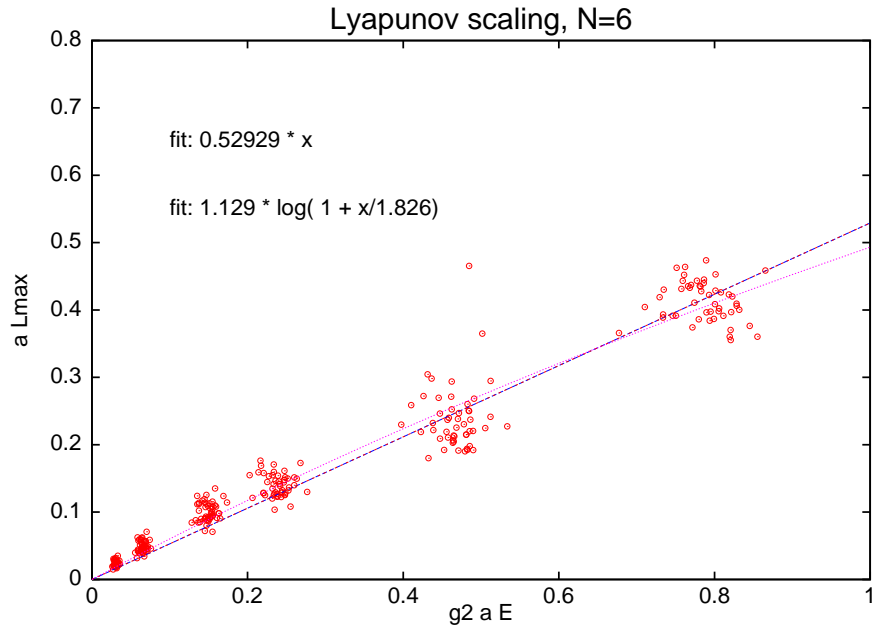}{85}
\caption{
 Scaling of short time Lyapunov exponents with energy 
 (before reaching the saturation in the distance of two
 initially adjacent configurations due to the compactness
 of the configuration space).
}
\end{figure}

\begin{figure}
\inplot{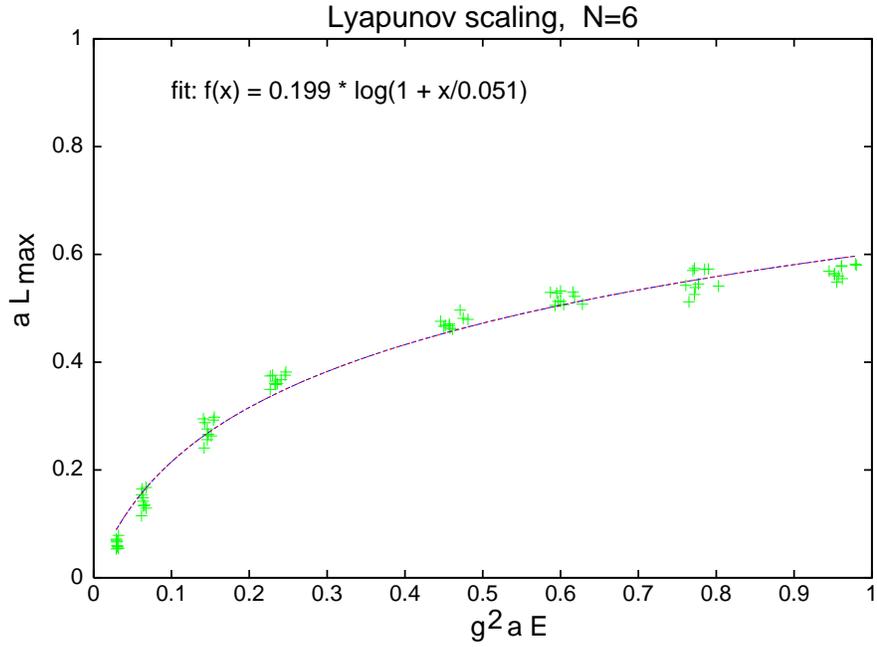}{85}
\caption{
  Scaling of Lyapunov exponents with energy for
  long time evolution of the monodromy matrix (after reaching
  distance saturation in the compact configuration space).
}
\end{figure}

\begin{figure}
\inplot{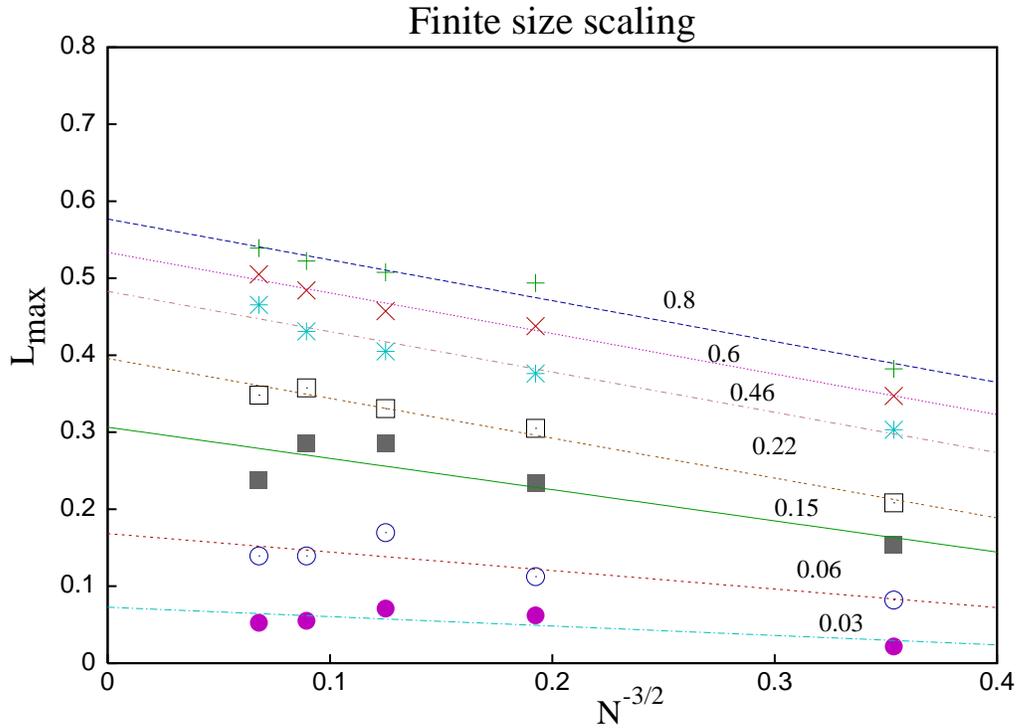}{85}
\caption{
 Finite size scaling of the maximal Lyapunov exponent.
 The straight line fits belong to different scaled energies
 (marked in the plot).
}
\end{figure}

\begin{figure}
\inplot{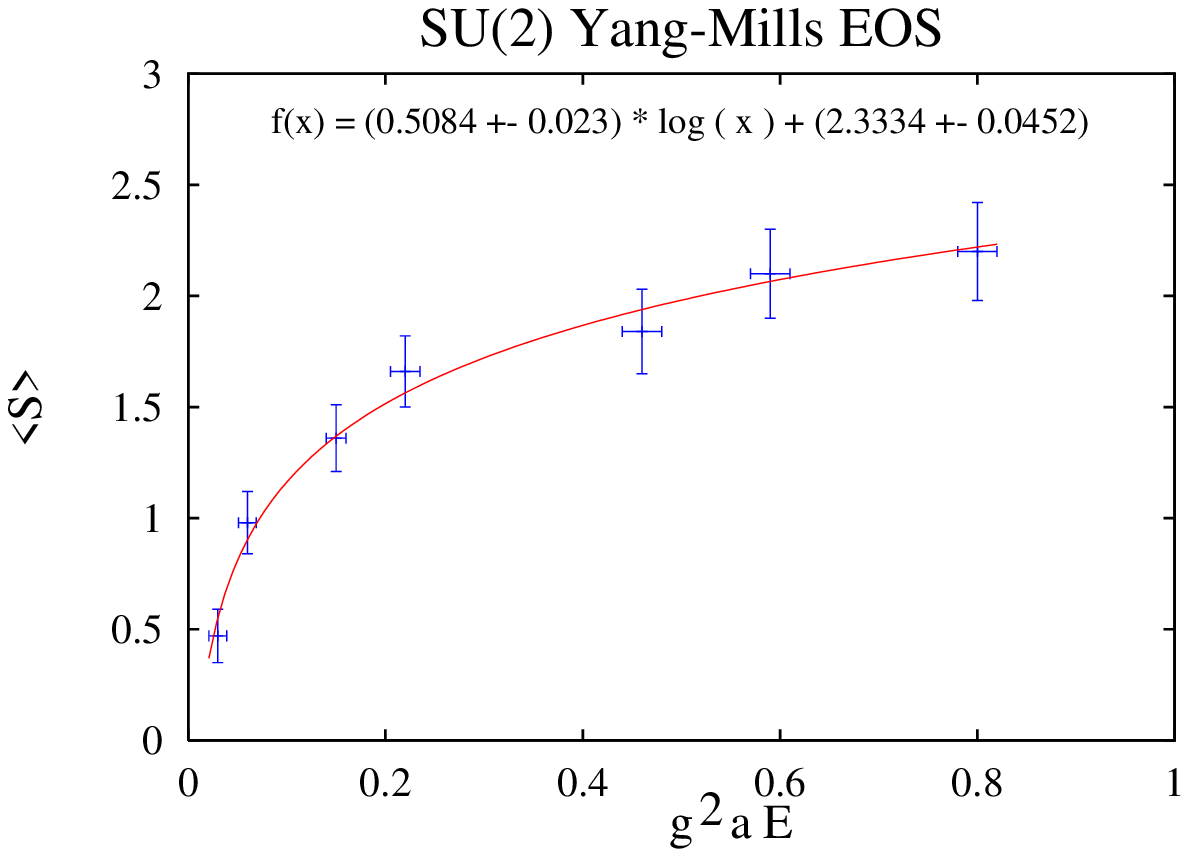}{100}
\caption{
 The normalized Kolmogorov-Sinai entropy. The best fit belongs to
 $\frac{1}{2} kT$ energy per degree of freedom, but a two
 phase structure also cannot be excluded (depletion hint).
}
\end{figure}



\begin{thebibliography}{99}
 \newcommand{\bi}{\bibitem}


\bi{blaiz1} J.-P. Blaizot, E. Iancu, A. Rebhan,
Approximately self-consitent resummations for the thermodynamics 
of the quark-gluon plasma.I. Entropy and density,
Phys. Rev. D 63:065003, 2001

\bi{blaiz2} J.-P. Blaizot, E. Iancu, A.Rebhan, The entropy of the QCD plasma, 
Phys. Rev. Lett. 83 (1999) 2906-2909

\bi{bt} T.S. Bir\'o, C. Gong, B. M\"uller,
 Lyapunov exponent and plasmon damping rate in non-Abelian
 gauge theories, 
 Phys. Rev. D 52 (1995) 1260-1266 

\bi{CG} C. Gong,
 Phys. Rev. D 49 (1994) 2642

\bi{bmat} B. M\"uller, A Trayanov,
 Phys. Rev. Lett 68 (1992)  3387

\bi{cgay} C. Gay, 
 Lyapunov exponent of Classical SU(3) Gauge Theory,
 Phys. Lett. B 298 (1993) 257-262


\bi{tsbagbmat}T. S. Bir\'o, C. Gong, B. M\"uller, A. Trayanov,
 Journal of Modern Physics C, Vol. 5, No. 1.(1994) 113-149


\bi{bffm}T.S. Bir\'o, \'A. F\"ul\"op, M. Feurstein, H. Markum,
  Investigation of Chaotic Dynamics of Lattice Gauge Configurations 
  Created by Monte Carlo Techniques,
  Conference Proceedings, "Strong and Electroweak Matter '97"
  Eger,(1997),  in World Scientific Publishing Co., (1997) p.304


\bi{BOOK} T.S. Bir\'o, S.G. Matinyan, B. M\"uller,
	Chaos and Gauge Field Theory, World Scientific 1995.


\bi{SEMI} H.Markum, R.Pulirsch, T.Wettig,
	Nonhermitian Random Matrix Theory and Lattice QCD
	with Chemical Potebtial, Phys.Rev.Lett. 83: 484, 1999

\bi{nrr} H.B. Nielsen, H.H. Rugh, S.E. Rugh, 
  Chaos and Scaling in Classical Non-Abelian Gauge Fields,  chao-dyn/9606013

\bi{nrr1} H.B. Nielsen, H.H. Rugh, S.E. Rugh, 
   Chaos, scaling and existence of a continuum limit in 
   classical non-Abelian lattice gauge theory,
   hep-th/9611128 

\bi{bm} B. M\"uller,
 Study of Chaos and Scaling in Classical SU(2) Gauge Theory,
 chao-dyn/9607001

\bi{hhlmm} U. Heinz, C. R. Hu, S. Leupold, S. G. Matinyan, B. M\"uller,
 Thermalization and Lyapunov exponens in the Yang-Mills Higgs Theory,
 Phys. Rev. D 55 (1997) 2464-2476

\bi{jbbmas} J. Bolte, B. M\"uller, A. Schafer, hep-lat/9906037
  Phys.Rev.D 61: 054506, 2000

\bi{s} D.R. Stump, Entropy of the SU(2) lattice gauge field,
Phys. Rev. D 36 (1987) 520-526



\end{thebibliography}
\end{document}